\documentclass[aps,
superscriptaddress,
showpacs,
preprintnumbers,
bibnotes,
amsmath,
amssymb,
twocolumn,
prl,
floatfix,
]{revtex4-2}
\usepackage[utf8]{inputenc}
\usepackage{array}
\usepackage[normalem]{ulem}
\newcolumntype{L}[1]{>{\raggedright\let\newline\\\arraybackslash\hspace{0pt}}m{#1}}
\newcolumntype{C}[1]{>{\centering\let\newline\\\arraybackslash\hspace{0pt}}m{#1}}
\newcolumntype{R}[1]{>{\raggedleft\let\newline\\\arraybackslash\hspace{0pt}}m{#1}}

\newcommand{\comment}[1]{}

\usepackage[caption=false]{subfig}

\usepackage{graphicx}
\usepackage{dcolumn}
\usepackage{bm}
\usepackage[usenames,dvipsnames]{color}
\usepackage{verbatim}
\usepackage{amsfonts}
\usepackage{longtable}
\usepackage{natbib}
\usepackage{hyperref}
\usepackage{dcolumn}

\newcolumntype{d}[1]{D{;}{.}{#1}}

\graphicspath{{figs/}}
\def\slashed#1{\kern+0.10em /\kern-0.50em #1}

\newcommand\bnl{Physics Department, Brookhaven National Laboratory, Upton, NY 11973, USA}

\newcommand\riken{RIKEN-BNL Research Center, Brookhaven National Laboratory, Upton, NY 11973, USA}
\newcommand\regensburg{Fakult\"at f\"ur Physik, Universit\"at Regensburg, Universit\"atsstra{\ss}e 31, 93040 Regensburg, Germany}
\newcommand\edinb{School of Physics and Astronomy, The University of Edinburgh, Edinburgh EH9 3FD, UK}
\newcommand\uconn{Physics Department, University of Connecticut, Storrs, CT 06269-3046, USA}
\newcommand\soton{School of Physics and Astronomy, University of Southampton,  Southampton SO17 1BJ, UK}

\newcommand\cern{CERN, Theoretical Physics Department, Geneva, Switzerland}
\newcommand\mib{Dipartimento di Fisica, Universit\`a di Milano-Bicocca, Piazza della Scienza 3, I-20126 Milano, Italy}
\newcommand\infn{INFN, Sezione di Milano-Bicocca, Piazza della Scienza 3, I-20126 Milano, Italy}

\newcommand{\llnl}{Nuclear and Chemical Sciences Division, Lawrence Livermore National Laboratory, Livermore, CA 94550, USA}

\begin{document}
\title{\boldmath The long-distance window of the hadronic vacuum polarization for the muon $g-2$}
\author{T.~Blum}\affiliation{\uconn}
\author{P.~A.~Boyle}\affiliation{\bnl}\affiliation{\edinb}
\author{M.~Bruno}\affiliation{\mib}\affiliation{\infn}
\author{B.~Chakraborty}\affiliation{\soton}
\author{F.~Erben}\affiliation{\cern}
\author{V.~G\"ulpers}\affiliation{\edinb}
\author{A.~Hackl}\affiliation{\regensburg}
\author{N.~Hermansson-Truedsson}\affiliation{\edinb}
\author{R.~C.~Hill}\affiliation{\edinb}
\author{T.~Izubuchi}\affiliation{\bnl}\affiliation{\riken}
\author{L.~Jin}\affiliation{\uconn}
\author{C.~Jung}\affiliation{\bnl}
\author{C.~Lehner}\thanks{Corresponding author}\email{christoph.lehner@ur.de}\affiliation{\regensburg}
\author{J.~McKeon}\affiliation{\soton}
\author{A.~S.~\surname{Meyer}}\affiliation{\llnl}
\author{M.~Tomii}\affiliation{\uconn}\affiliation{\riken}
\author{J.~T.~Tsang}\affiliation{\cern}
\author{X.-Y.~Tuo}\affiliation{\bnl}

\collaboration{RBC and UKQCD Collaborations}
\noaffiliation

\date{\today}


\preprint{CERN-TH-2024-182}


\begin{abstract}
  We provide the first ab-initio calculation of the Euclidean
long-distance window of the isospin symmetric light-quark connected
contribution to the hadronic vacuum polarization for the muon $g-2$ and find
$a_\mu^{\rm LD,iso,conn,ud} = 411.4(4.3)(2.4) \times 10^{-10}$.  We
also provide the currently most precise calculation of the
total isospin symmetric light-quark connected contribution,
$a_\mu^{\rm iso,conn,ud} = 666.2(4.3)(2.5) \times 10^{-10}$, which
is more than 4$\sigma$ larger compared to the data-driven estimates
of Boito et al. 2022 and 1.7$\sigma$ larger compared to the lattice
QCD result of BMW20.
\end{abstract}

\maketitle

%


\section{Introduction}
The relative deviation of the muon's Land\'e factor $g_\mu$ from
Dirac's relativistic quantum mechanics result, $a_\mu = (g_\mu -
2)/2$, also called the anomalous magnetic moment of the muon, is one of
the most precisely determined quantities in particle physics.  It is
sensitive to virtual contributions of particles which may be out of
reach of direct production in high-energy experiments and it therefore
plays an important role in constraining new physics.  Substantial
efforts have been undertaken at Fermilab (E989) and are planned at
J-PARC (E34) \cite{Abe:2019thb} in order to further improve the
precision of the experimental determination.  In 2021 the Fermilab
experiment released first results \cite{Muong-2:2021ojo} which
confirmed the previously best result obtained by the BNL E821
experiment \cite{Bennett:2006fi} and reduced the experimental
uncertainty from 0.54~ppm to 0.46~ppm. Subsequently they released results for runs II and III in 2023 which yielded an uncertainty of 0.2 ppm~\cite{Muong-2:2023cdq}. In 2025 the Fermilab
experiment aims to release their final results pushing the uncertainty
down to approximately 0.14~ppm \cite{Carey:2009zzb}.

Matching the precision of this spectacular experimental result in a theory
calculation of the Standard Model (SM) contribution to $a_\mu$ is a substantial
challenge and is currently a work in progress.  In 2020 the Muon g-2 Theory
Initiative published a whitepaper \cite{Aoyama:2020ynm,Aoyama:2012wk,Aoyama:2019ryr,Czarnecki:2002nt,Gnendiger:2013pva,Davier:2017zfy,Keshavarzi:2018mgv,Colangelo:2018mtw,Hoferichter:2019mqg,Davier:2019can,Keshavarzi:2019abf,Kurz:2014wya,Melnikov:2003xd,Masjuan:2017tvw,Colangelo:2017fiz,Hoferichter:2018kwz,Gerardin:2019vio,Bijnens:2019ghy,Colangelo:2019uex,Blum:2019ugy,Colangelo:2014qya} which indicated a more than $4\sigma$ tension
of the experimental result with the SM.  This result relied on a data-driven
estimate of the hadronic vacuum polarization (HVP) contribution based on $e^+e^-\to$
hadron experimental data.  While existing tensions between experimental data sets
had been taken into account in 2020 by inflating the uncertainties appropriately,
the recent result by CMD-3 \cite{CMD-3:2023rfe,CMD-3:2023alj} increases the tensions
to a degree that currently appears to make a data-driven high-precision evaluation of the HVP not feasible.

At the same time, lattice methodology is maturing and is on track to
allow for a complete ab-initio theory determination that soon may
match the Fermilab E989 target precision.  In this context, it is now
common practice to separate the total HVP contribution into the
Euclidean windows introduced in our previous work \cite{RBC:2018dos},
which separate three different regions (short-distance,
intermediate-distance, long-distance) that exactly sum up to the
total HVP.  Each region has its own dominant challenges that can best be
addressed by targeted calculations optimized separately for each region.
 
The short-distance region suffers from large discretization errors and
very fine lattices are needed.  The intermediate-distance region has
moderate uncertainties.  The
long-distance region suffers from large statistical and
finite-size errors.  For lattice calculations using rooted staggered
quarks additional challenges for the continuum limit of the
long-distance contribution exist, see,
e.g. Ref.~\cite{Borsanyi:2020mff}.  

By now there is very good
agreement among several lattice collaborations on the short-distance
and intermediate-distance windows
\cite{RBC:2023pvn,Aubin:2019usy,Giusti:2021dvd,Borsanyi:2020mff,Lehner:2020crt,Aubin:2022hgm,Wang:2022lkq,Ce:2022kxy,Alexandrou:2022amy,Bazavov:2023has,Giusti:2021dvd,Alexandrou:2022amy,Boccaletti:2024guq,Kuberski:2024bcj,Spiegel:2024dec}.
The same consolidation at high precision is also needed for the long-distance
region, and the current paper which focuses on the dominant light-quark connected contribution is a first step in this direction.  The results reported in the following are unchanged compared to our
unblinding presentation at Lattice 2024 \cite{Lattice24Talk}.

In future work, we will improve our previous results \cite{RBC:2018dos} on the
quark disconnected contributions, strong-isospin-breaking (SIB)
and QED corrections and improve our lattice spacing determinations to complete the calculation of $a_\mu$ at the next
precision frontier.  We note that a combination of lattice and data-driven
methodology is also an interesting approach as suggested in Ref.~\cite{RBC:2018dos}
and recently demonstrated in Ref.~\cite{Boccaletti:2024guq}.
  
\section{Methodology}\label{sec:methodology}
We address the particular challenges of the long-distance window by
building on ideas of the improved bounding method \cite{Bruno:2019nzm}
and finite-volume exclusive state reconstruction
\cite{DellaMorte:2017khn}.  These methods are expressed in terms
of the time-momentum representation
\cite{Bernecker:2011gh}
\begin{align}
  a^{\rm HVP~LO}_\mu &= \sum_{t=0}^\infty w_t C(t)
  \label{eq:tmr}
\end{align}
with $C(t) = \frac13 \sum_{\vec{x}}\sum_{j=0,1,2} \langle J_j(\vec{x},t) J_j(0) \rangle$, vector current 
$J_\mu(x) = i\sum_f Q_f \overline{\Psi}_f(x) \gamma_\mu \Psi_f(x)$ with fractional electric charge $Q_f$, and sum over quark flavors $f$.  The correlator $C(t)$ at zero temperature admits a spectral representation
\begin{align}
  C(t) = \frac13 \sum_{j=0,1,2}\sum_{n} \vert \langle n \vert \hat J_j \vert 0 \rangle \vert^2 e^{-E_n t}
\end{align}
with zero-momentum vector operator $\hat{J}_j$, Hamiltonian eigenstate $\vert n \rangle$, and energy $E_n$
for the discrete finite-volume spectrum.  The weights $w_t$ can be
discretized in different ways.  We use the two approaches described in
Ref.~\cite{RBC:2023pvn}.  We drop the HVP~LO label in the following
and separate the window contributions as in Ref.~\cite{RBC:2018dos}
into $a_\mu = a_\mu^{\rm SD} + a_\mu^{\rm W} + a_\mu^{\rm LD}$.  We
have $a_\mu^{\rm SD}(t_0,\Delta) = \sum_{t=0}^\infty C(t) w_t [1 -
  \Theta(t,t_0,\Delta)]$, $a_\mu^{\rm W}(t_0,t_1,\Delta) =
\sum_{t=0}^\infty C(t) w_t [ \Theta(t,t_0,\Delta) -
  \Theta(t,t_1,\Delta) ]$, and $a_\mu^{\rm LD}(t_1,\Delta) =
\sum_{t=0}^\infty C(t) w_t \Theta(t,t_1,\Delta)$ with
$\Theta(t,t',\Delta) = \left[1 + \tanh\left[ (t-t') / \Delta
    \right]\right]/2$.  We select $t_0=0.4$ fm, $t_1=1.0$ fm, and
$\Delta=0.15$ fm as suggested in Ref.~\cite{RBC:2018dos}.

The correlators $C(t)$ are computed with a hierarchical approximation
scheme \cite{Blum:2012uh,Shintani:2014vja,RBC:2018dos} using
locally-coherent low-modes with exact eigenvalues
\cite{Clark:2017wom}.  In addition, as described in
Ref.~\cite{Bruno:2019nzm}, we use exact distillation
\cite{HadronSpectrum:2009krc,Bruno:2023pde} and have made our code
publicly available \cite{GPT}.

\begin{table}
\begin{ruledtabular}
  \begin{tabular}{l|lllllllll}
  ID & $a^{-1}$/GeV &  $L^3 \times T \times L_s/a^5$ & $m_\pi$/MeV & $m_K$/MeV & $m_\pi L$  \\\hline
  96I & $2.6920(67)$ &  $96^3 \times 192 \times 12$  &  $131.29(66)$ & $484.5(2.3)$& 4.7 \\
  64I & $2.3549(49)$ &  $64^3 \times 128 \times 12$ &  $138.98(43)$ & $507.5(1.5)$ & 3.8 \\
  48I & $1.7312(28)$ &  $48^3 \times 96 \times 24$ & 
  $139.32(30)$ & $499.44(88)$ & 3.9 \\
   C & $1.7312(28)$ &  $64^3 \times 96 \times 24$ &  $139.32(30)$ & $499.44(88)$ & 5.2 \\\hline
  4   & $1.7312(28)$ &  $24^3 \times 48 \times 24$  &  $274.8(2.5)$ & $530.1(3.1)$ & 3.8  \\
  D   & $1.7312(28)$&  $32^3 \times 48 \times 24$  &  $274.8(2.5)$ & $530.1(3.1)$ & 3.8  \\
  1   & $1.7312(28)$ &  $32^3 \times 64 \times 24$  &  $208.1(1.1)$ & $514.0(1.8)$ & 3.8  \\
  3   &$1.7312(28)$ &  $32^3 \times 64 \times 24$  &  $211.3(2.3)$ & $603.8(6.1)$ & 3.9  \\
  9   & $2.3549(49)$ &  $32^3 \times 64 \times 12$  &  $278.9(0.6)$ & $531.2(0.7)$ & 3.8  \\
 L   & $2.3549(49)$ &  $64^3 \times 128 \times 12$  &  $278.9(0.6)$ & $531.2(0.7)$ & 7.6  \\
  \end{tabular}
\end{ruledtabular}
\caption{\label{tab:ex} List of $N_f=2+1$ ensembles with parameters determined in the RBC/UKQCD18 isospin symmetric world defined in Eq.~\ref{eqn:rbcworld}.  The ensembles have
  Iwasaki gauge action and M\"obius \cite{Brower:2012vk} domain-wall \cite{Shamir:1993zy,Furman:1994ky} fermion sea quarks with $b=1.5$ and $c=0.5$.  The parameters $b$ and $c$ are defined in Ref.~\cite{RBC:2014ntl}.  The scripts generating the new ensembles
are publicly available \cite{GPT}.}
\end{table}

In Tab.~\ref{tab:ex} we provide a list of all lattice gauge ensembles
used in this study.  For the physical pion mass ensembles with $m_\pi
L \approx 5$ (96I and C) we use 200 Laplace eigenmodes and for all other ensembles we use 60.  We use an operator basis including a
local vector current, a distillation-smeared vector current, and
two-pion operators up
to a relative momentum of $p=(2,0,0)(2\pi/L)$
for the case of 60 Laplace eigenmodes and $p=(2,2,0)(2\pi/L)$ for the
case of 200 Laplace eigenmodes.  All operators are in the $T_1^u$, $I=1$ irreducible representation since we focus on the dominant light-quark connected contribution.

\begin{figure}
\includegraphics[width=\linewidth]{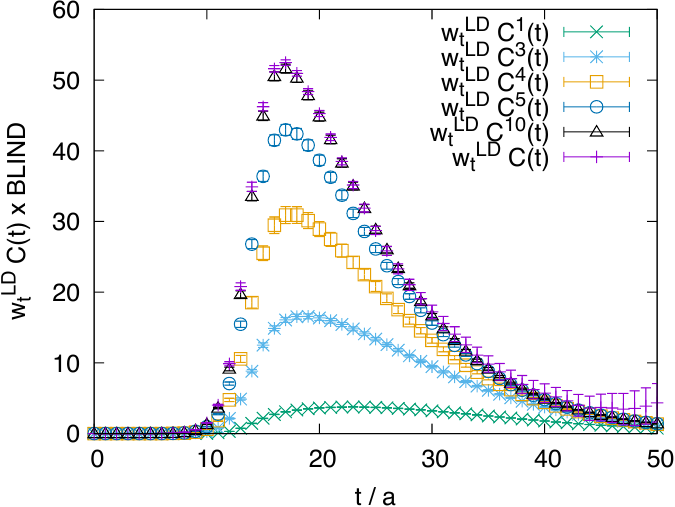}
\caption{\label{fig:reconst} Reconstruction of the lowest $N$ states
  $C^N(t)$ compared to the inclusive $C(t)$ for the 96I ensemble with
  $w_t^{\rm LD}=w_t \Theta(t,t_1,\Delta)$.  The exponential growth of
  statistical noise in $C(t)$ is absent in the reconstruction.}
\end{figure}

In Fig.~\ref{fig:reconst}, we
demonstrate the reconstruction of the first $N$ states via $C^N(t) =
(1/3)\sum_{j=0,1,2} \sum_{n=1}^{N} \vert \langle n \vert \hat J_j \vert 0 \rangle \vert^2 e^{-E_n t}$
compared to the full $C(t)$.  The figure shows the result of using
only distillation smeared operators (vector and two-pion) in a
generalized eigenvalue problem (GEVP) study \cite{Luscher:1990ck,Blossier:2009kd} to find the $E_n$ and
optimal operators to project to a given state $\vert n\rangle$.  This operator is
contracted with the local vector current to obtain the overlap
factor $\vert \langle n \vert \hat J_j \vert 0 \rangle \vert^2$.
The reconstruction $C^N(t)$ is used at large Euclidean times $t$,
where $C^N(t) = C(t)$ within statistical uncertainties.

As in our previous work \cite{RBC:2023pvn}, we perform the analysis in a blinded manner with
five analysis groups A-E.  Four analysis groups A-D have conducted a full analysis, while one group (E)
has focused on cross-checks.  Each analysis group received the correlator data with a blinding factor
applied to each insertion of a local vector current.  The blinding factor was unique to each group and generated
using a hash function based on the group's name (A-E).  The blinding factors were applied by a script, and no
one in the collaboration saw the blinding factors themselves.  The process was managed
by one of the authors (CL).  Once the analysis groups were ready, relative unblinding meetings between
two analysis groups were organized in which the respective methods were scrutinized.  After conclusion of this
process the relative blinding factor was removed between the participating groups.  After the relative unblinding,
the collaboration agreed to the RBC/UKQCD24 prescription for the analysis that will be described in detail in the following.  This procedure was then applied to a final analysis before the absolute blinding was removed in a joint meeting on
July 19, 2024.  The cross-checks of group E were conducted prior to the unblinding and focused on results that were not affected by the blinding factor such as the ratio of $C^N(t)$ to $C(t)$ for which the blinding factor drops out, as well as checks of individual $E_n$ that are also not affected by the blinding.  Additional details
on the cross-checks and on the various methods studied by the individual groups are provided as supplemental
material to this letter.

The calculation of $C(t)$ is organized as an expansion around an
isospin-symmetric point
\cite{deDivitiis:2013xla,Boyle:2017gzv,RBC:2018dos,Giusti:2019xct,DiCarlo:2019thl,Boyle:2022lsi}.
In this work, we provide results for two choices of the expansion point, following our
earlier work \cite{RBC:2023pvn}.  The
first choice is the RBC/UKQCD18 world defined by
\begin{align}\label{eqn:rbcworld}
m_\pi&=0.135~\text{GeV}\,, &
m_K&=0.4957~\text{GeV}\,, \notag\\
m_\Omega &= 1.67225~\text{GeV} \,,
\end{align}
consistent with Ref.~\cite{RBC:2018dos}.
We also consider a second choice
\begin{align}\label{eqn:bmwworld}
  m_\pi&=0.13497~\text{GeV}\,, &
  m_{ss*}&=0.6898~\text{GeV}\,, \notag\\
  w_0 &=0.17236~\text{fm} \,,
\end{align}
which we label as the BMW20 world \cite{Borsanyi:2020mff}.  We define
$m_{ss*}$ as the ground-state energy of the quark-connected
pseudoscalar $\bar{s}s$ meson two-point function.  To avoid an
unnecessary inflation of uncertainties when comparing
isospin-symmetric lattice results, we define the above values without
uncertainties.  The experimental uncertainties of the physical hadron
spectrum enter when QED and SIB corrections are included.

\begin{figure}
\includegraphics[width=\linewidth]{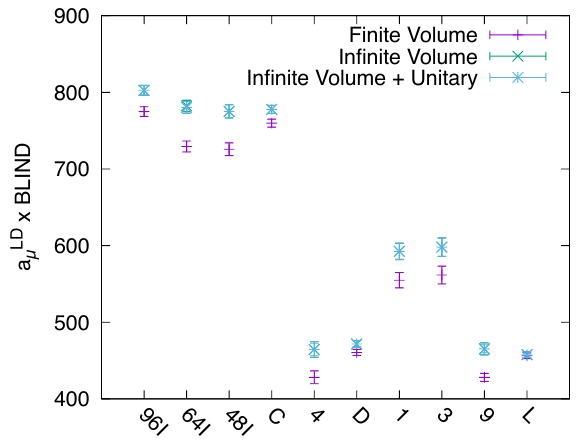}
\caption{\label{fig:fvresults} Results for the ensembles listed in
  Tab.~\ref{tab:ex} with and without finite-volume corrections
  applied.  For ensemble 64I a small correction from the
  partially-quenched point to the unitary point is added as
  well, see Ref.~\cite{RBC:2023pvn}.  }
\end{figure}

Finite-volume corrections are applied using the Hansen-Patella
formalism \cite{Hansen:2019rbh,Hansen:2020whp} in the monopole
approximation.  As can be seen in Fig.~\ref{fig:fvresults}, after
applying finite-volume corrections the results for ensembles that
only differ by the lattice volume (4 and D, 9 and L, 48I and C)
agree within uncertainties.  We provide a detailed study of the agreement
as a function of Euclidean time in the supplemental material.  It is
noteworthy that for our largest volume with $m_\pi L=7.6$ (L), the
finite-volume corrections are smaller than the quoted statistical
uncertainties.

\section{Analysis and results}
In the following, we describe the RBC/UKQCD24 prescription for
determining the light-quark connected contribution to the
long-distance window in the isospin symmetric limit, $a_\mu^{\rm
  LD,iso,conn,ud}$. 

\begin{figure}
\includegraphics[width=\linewidth]{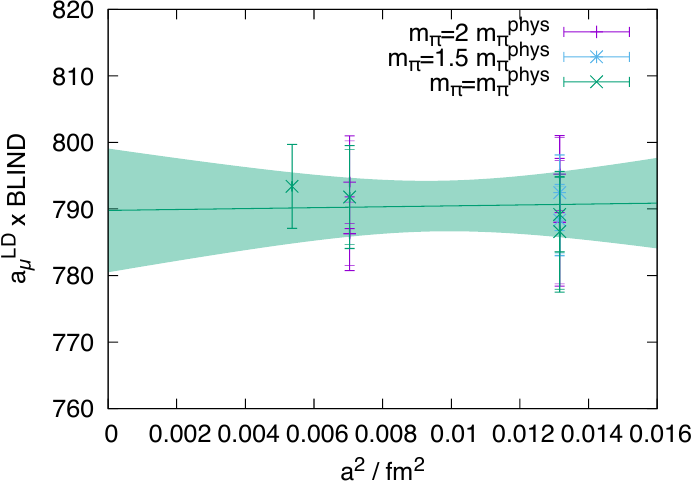}
\caption{\label{fig:mult} Fit result of $f_*$ with
  $Z_V^\pi$ and without setting $f_3=0$.  The data is shown for all ensembles as a function of $a^2$ after subtracting the fit function without the $f_0 f_1 a^2$ term.}
\end{figure}

For each ensemble in Tab.~\ref{tab:ex}, the long-distance part of
$C(t)$ is replaced by $C^{N}(t)$ for sufficiently large times such
that they agree within statistical uncertainties.  For the ensembles
with 200 Laplace modes $N=10$ and for all other ensembles $N=5$.  In all
cases this allows for a spectral reconstruction beyond the peak of the
rho resonance.  Finite-volume corrections are applied as shown in
Fig.~\ref{fig:fvresults}.  These data points are fit jointly to
several fit functions.  We study both an additive and a multiplicative
combination of discretization and mass-mistuning effects by varying
between
\begin{align}
  f_{+} &= f_0 + f_1 a^2 + f_2 (w_0 m_\pi - (w_0 m_\pi)_{\rm phys}) \notag\\
  &\quad + f_3 (w_0 m_\pi - (w_0 m_\pi)_{\rm phys})^2 \notag\\
  &\quad + f_4 (w_0 m_{ss*} - (w_0 m_{ss*})_{\rm phys})
\end{align}
and
\begin{align}
  f_{*} &= f_0(1 + f_1 a^2)(1 + f_2 (w_0 m_\pi - (w_0 m_\pi)_{\rm phys}) \notag\\
  &\quad + f_3 (w_0 m_\pi - (w_0 m_\pi)_{\rm phys})^2 \notag\\
  &\quad + f_4 (w_0 m_{ss*} - (w_0 m_{ss*})_{\rm phys})) \,.
\end{align}
The functional forms as given apply to the BMW20 world and the
dimensionless ratios $w_0 m_\pi$ and $w_0 m_{ss*}$ have to be replaced
with $m_\pi / m_\Omega$ and $m_K / m_\Omega$ for the RBC/UKQCD18
world.  For both fit functions, we also study versions with $f_3=0$.
These four fit forms are then applied to the data renormalized with
two different choices for the local vector current renormalization
constant: $Z_V^\pi$ and $Z_V^\star$.  The former is defined by the
pion charge, the latter by the ratio of local-conserved to local-local
correlators at a distance of 1 fm.  This results in 8 fits that are
then combined in a model average.  All fit forms have acceptable
$p$-value and the results are consistent between using the Akaike
information criterion (AIC) \cite{Akaike:1974vps}, a simple $\chi^2$
weight, and a flat weight of all models.  We provide individual
results in the supplemental material.  We also studied more divergent
chiral dependencies, however, since our analysis is dominated by
four ensembles at physical pion mass, such variations have little impact on
the fit results.  In Fig.~\ref{fig:mult}, we show the fit result of $f_*$
with $Z_V^\pi$ and without setting $f_3=0$.  We emphasize that the extrapolation
to the continuum limit is within the statistical uncertainties
of the finest data point.

\begin{figure}
  \includegraphics[page=1,width=\linewidth]{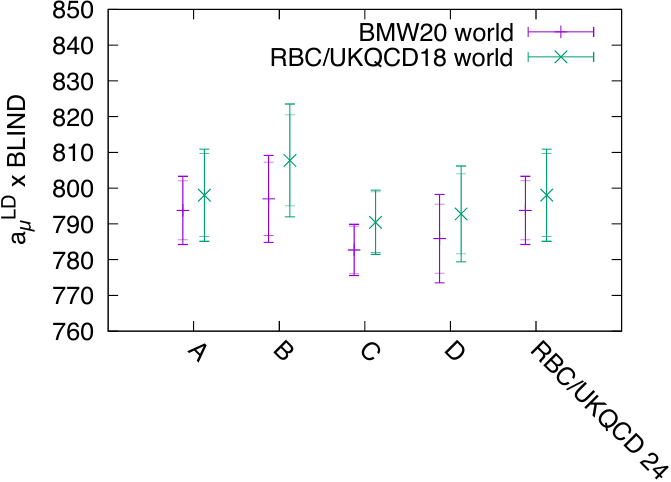}
  \caption{\label{fig:compgroups}Results obtained
    by the different analysis groups and the resulting RBC/UKQCD24 prescription.
}
\end{figure}
In Fig.~\ref{fig:compgroups}, we compare the results obtained by the
different analysis groups to the RBC/UKQCD24 prescription.  We
observed good agreement prior to the absolute unblinding and have
identified the reasons for the residual variations.  We note that group
D only took the continuum limit of physical pion mass ensembles.
Groups A and B also verified the consistency of the continuum limits
with and without ensembles 9 and L.  The lattice spacing uncertainty
due to our more limited knowledge of the $\Omega^-$ mass is
responsible for the larger errors in RBC/UKQCD18 world.  Work on a
more precise determination of $m_\Omega$ is in progress.  We observe
that RBC/UKQCD18 and BMW20 worlds are consistent at the current precision.

\begin{figure}[tb]
\includegraphics[width=\linewidth]{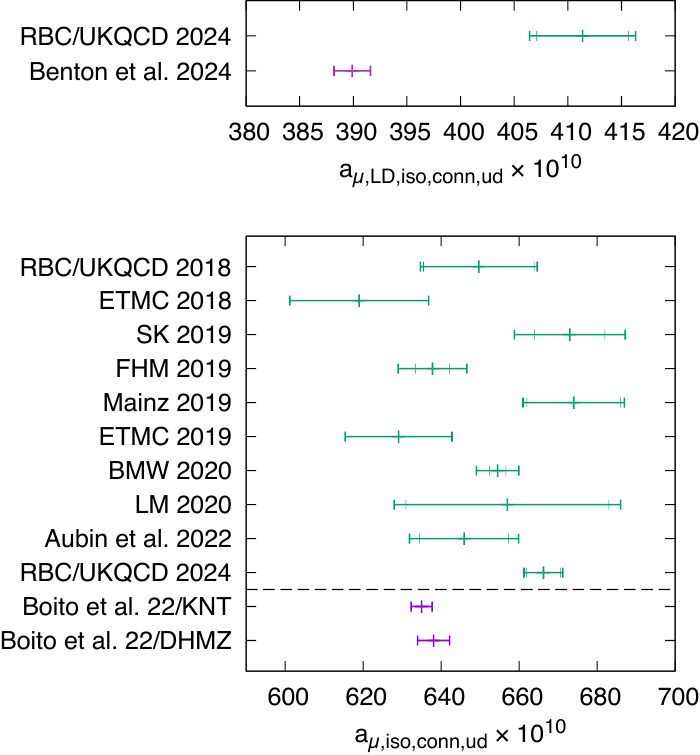}
\caption{\label{fig:comp} This work compared to the literature:
  Benton et al. 2024 \cite{Benton2024},
  RBC/UKQCD 2018 \cite{RBC:2018dos},
  ETMC 2018 \cite{Giusti:2018mdh},
  SK 2019 \cite{Shintani:2019wai},
  FHM 2019 \cite{FermilabLattice:2019ugu},
  Mainz 2019 \cite{Gerardin:2019rua},
  ETMC 2019 \cite{Giusti:2019hkz},
  BMW 2020 \cite{Borsanyi:2020mff},
  LM 2020 \cite{Lehner:2020crt},
  Aubin et al. 2022 \cite{Aubin:2022hgm}, and
  Boito et al. 2022/KNT and Boito et al. 2022/DHMZ \cite{Boito:2022dry}.
}
\end{figure}

Our final results are
\begin{align}
  a_\mu^{\rm LD,iso,conn,ud} &= 411.4(4.3)(2.4) \times 10^{-10} \,, \notag\\
  a_\mu^{\rm iso,conn,ud} &= 666.2(4.3)(2.5) \times 10^{-10}
\end{align}
in the BMW20 world and
\begin{align}
  a_\mu^{\rm LD,iso,conn,ud} &= 413.6(6.0)(2.9) \times 10^{-10} \,, \notag\\
  a_\mu^{\rm iso,conn,ud} &= 668.7(6.1)(2.9) \times 10^{-10}
\end{align}
in the RBC/UKQCD18 world, where the first error is statistical and the second systematic.  The total isospin symmetric results are
obtained by adding our previous short-distance and intermediate-distance
results \cite{RBC:2023pvn}.

In Fig.~\ref{fig:comp}, we compare our results in the BMW20 world to
the literature \footnote{We note that after the unblinding of our result at Lattice 2024 in July of 2024, at the KEK workshop of the Muon g-2 Theory Initiative in September of 2024
unblinded results for $a_\mu^{\rm LD,iso,conn,ud}$ and $a_\mu^{\rm iso,conn,ud}$ were presented by the Mainz collaboration.  At the point of finalizing this manuscript the Mainz result had not yet appeared in the literature.}.  Our result for $a_\mu^{\rm iso,conn,ud}$ is more than
4$\sigma$ larger compared to the data-driven estimates by Boito, {\it et
al}. 2022 \cite{Boito:2022dry} which were obtained based on the data sets that entered the Theory Initiative
whitepaper \cite{Aoyama:2020ynm} prior to the release
of the CMD-3 data.  This observed shift with respect to the data-driven estimate is consistent with the size of the tension between experiment and theory for the muon
$g-2$ quoted in the 2020 whitepaper of the Muon g-2 Theory Initiative
\cite{Aoyama:2020ynm}.  Finally, we note that our result is also 1.7$\sigma$ larger compared to the
lattice QCD result of BMW20 \cite{Borsanyi:2020mff}.



\section{Conclusions and Outlook}\label{sec:conclude}
In this work we compute the long-distance Euclidean window of the
hadronic vacuum polarization for light quarks in the isospin
symmetric limit.  This calculation is particularly challenging and 
dominates the total uncertainty of a complete high-precision
$a_\mu^{\rm HVP~LO}$ result obtained from first-principles lattice QCD
methods.  All calculations were performed in a blinded
manner.  We find a large tension with the data-driven approach \cite{Boito:2022dry} based on data sets that were also used in Ref.~\cite{Aoyama:2020ynm} but also a smaller tension with
the BMW20 result \cite{Borsanyi:2020mff}.  More work is needed to
complete an ab-initio calculation matching the Fermilab E989 target
precision.  We are currently improving our previous estimates for the
quark-disconnected contributions and the QED and SIB corrections
including diagrams beyond the electro-quenched approximation.  We
expect to match the FNAL E989 target precision upon
completion of our HVP program.

\section{Acknowledgments}
We thank our colleagues of the RBC and UKQCD collaborations for many valuable discussions and joint efforts over the years.
The authors gratefully acknowledge the Gauss Centre for Supercomputing
e.V. (www.gauss-centre.eu) for funding this project by providing
computing time on the GCS Supercomputer JUWELS at Jülich
Supercomputing Centre (JSC).  We acknowledge the EuroHPC Joint
Undertaking for awarding this project access to the EuroHPC
supercomputer LUMI, hosted by CSC (Finland) and the LUMI consortium
through a EuroHPC Extreme Scale Access call as well as the EuroHPC
supercomputer LEONARDO, hosted by CINECA (Italy).  An award of
computer time was provided by the ASCR Leadership Computing Challenge
(ALCC) and Innovative and Novel Computational Impact on Theory and
Experiment (INCITE) programs. This research used resources of the
Argonne Leadership Computing Facility, which is a DOE Office of
Science User Facility supported under contract DE-AC02-06CH11357. This
research also used resources of the Oak Ridge Leadership Computing
Facility, which is a DOE Office of Science User Facility supported
under Contract DE-AC05-00OR22725.  This research used resources of the
National Energy Research Scientific Computing Center (NERSC), a
U.S. Department of Energy Office of Science User Facility located at
Lawrence Berkeley National Laboratory, operated under Contract
No.~DE-AC02-05CH11231 using NERSC award NESAP m1759 for 2020. This
work used the DiRAC Blue Gene Q Shared Petaflop system at the
University of Edinburgh, operated by the Edinburgh Parallel Computing
Centre on behalf of the STFC DiRAC HPC Facility
(www.dirac.ac.uk). This equipment was funded by BIS National
E-infrastructure capital grant ST/K000411/1, STFC capital grant
ST/H008845/1, and STFC DiRAC Operations grants ST/K005804/1 and
ST/K005790/1. DiRAC is part of the National E-Infrastructure.  We
gratefully acknowledge disk and tape storage provided by USQCD and by
the University of Regensburg with support from the DFG.
The lattice data analyzed in this project was generated using GPT
\cite{GPT}, Grid \cite{GRID}, and CPS \cite{CPS} and analyzed, in
part, using pyobs \cite{PYOBS}.
TB is supported by the US DOE under grant DE-SC0010339.  PB, TI, and CJ
were supported in part by US DOE Contract DESC0012704(BNL) and
the Scientific Discovery
through Advanced Computing (SciDAC) program LAB 22-2580.  At the beginning of the project, MB was supported by the national program for young researchers “Rita Levi Montalcini”.
MB is (partially) supported by ICSC - Centro Nazionale di Ricerca in High Performance Computing, Big Data and Quantum Computing, funded by European Union – NextGenerationEU.
Research of BC at the University of Southampton has been supported by the following research fellowship and grants - Leverhulme Trust (ECF-2019-223 G100820), STFC (Grant no. ST/X000583/1), STFC (Grant no. ST/W006251/1), and EPSRC (Grant no. EP/W032635/1).
FE has received funding from the European Union’s Horizon Europe research and innovation programme under the Marie Sk\l{}odowska-Curie grant agreement No. 101106913.
VG is supported in part by UK STFC grant ST/P000630/1, ST/T000600/1 and ST/X000494/1.
AH is supported by the Hans-B\"ockler-Stiftung.
 NHT is funded by the UK Research and Innovation, Engineering and Physical Sciences Research Council, grant number EP/X021971/1. 
 RH is supported by UK STFC Grant No. ST/P000630/1. 
 TI is also
supported by the Department of Energy, Laboratory Directed Research
and Development (LDRD No. 23-051) of BNL and RIKEN BNL Research
Center.  LJ and MT acknowledge the support of DOE Office of Science Early
Career Award DE-SC0021147 and DOE grant DE-SC0010339.  
ASM is supported in part by Lawrence Livermore National Security, LLC \#DE-AC52-07NA27344, Neutrino Theory Network Program Grant \#DE-AC02-07CHI11359, and the U.S. Department of Energy Award \#DE-SC0020250.  XYT has been supported by US DOE Contract DESC0012704(BNL).

\bibliography{references}

\clearpage
\appendix
\section*{Supplemental Material}

\section{Test of the finite-volume corrections}
\begin{figure*}
\includegraphics[page=3,width=0.48\linewidth]{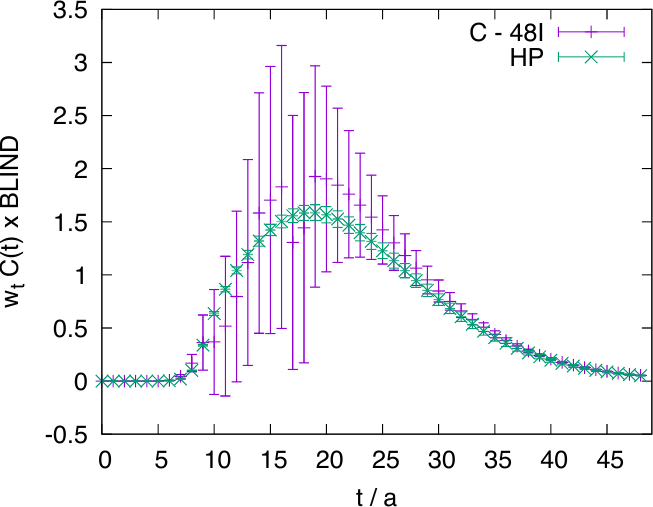}
\includegraphics[page=4,width=0.48\linewidth]{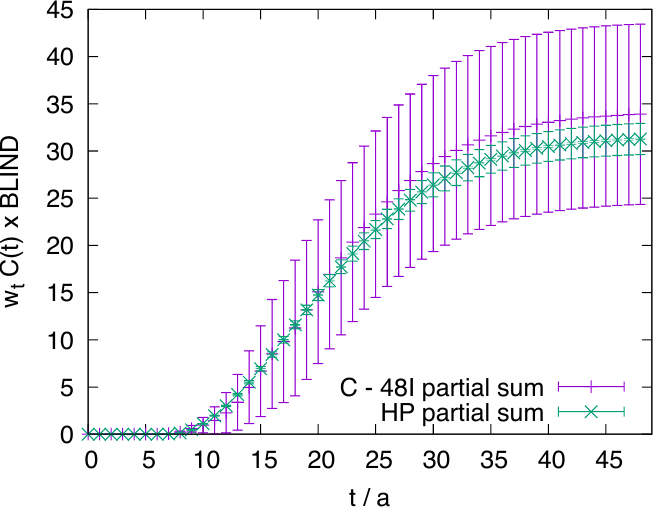}
\includegraphics[page=7,width=0.48\linewidth]{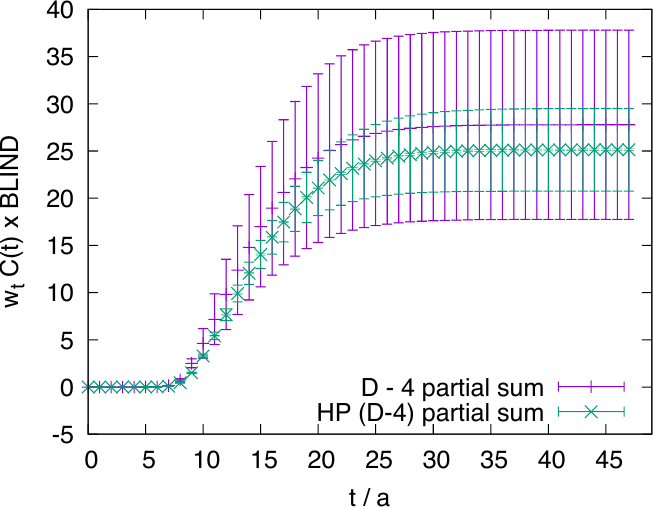}
\includegraphics[page=8,width=0.48\linewidth]{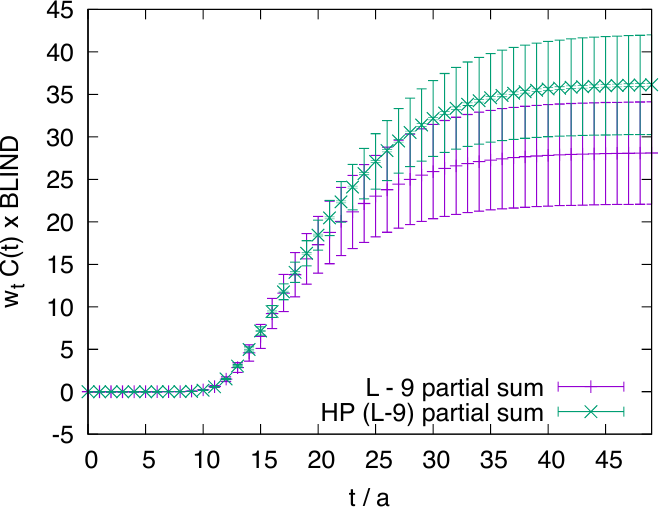}
\caption{\label{fig:fvcorrtest} We show the difference of ensemble pairs in the time-momentum representation which only differ by the lattice volume and compare the result to the Hansen-Patella formalism in the monopole ansatz.
In the top left panel, we show the integrand for ensembles C and 48I at physical pion mass and in the top right panel, we show the partial sum $\sum_{t'=0}^t w_{t'} C(t')$ for the same ensembles.  The bottom row shows the partial sums of the integrands for ensembles with pion mass approximately equal to 280 MeV from $m_\pi L\approx 4$ to $m_\pi L \approx 8$.
}
\end{figure*}
In Fig.~\ref{fig:fvcorrtest}, we provide additional details about the
excellent agreement of the Hansen-Patella
formalism \cite{Hansen:2019rbh,Hansen:2020whp} with the observed
behavior in the lattice data for both physical pion mass and up to
$m_\pi \approx 280$ MeV.  We use the monopole ansatz and vary the
resonance mass parameter from $m_\rho=727$ to $770$ MeV to generate
the uncertainty estimates.  Remarkably, the lattice data agrees
for fixed Euclidean time $t$ in all studied cases very
well with the Hansen-Patella results.  Our study extends from $m_\pi
L \approx 4$ to $m_\pi L \approx 8$ for which the finite-volume
corrections are smaller than the statistical uncertainties of the
lattice result.

\section{Distinct features of individual analysis groups}
\begin{figure*}
\includegraphics[width=0.48\linewidth]{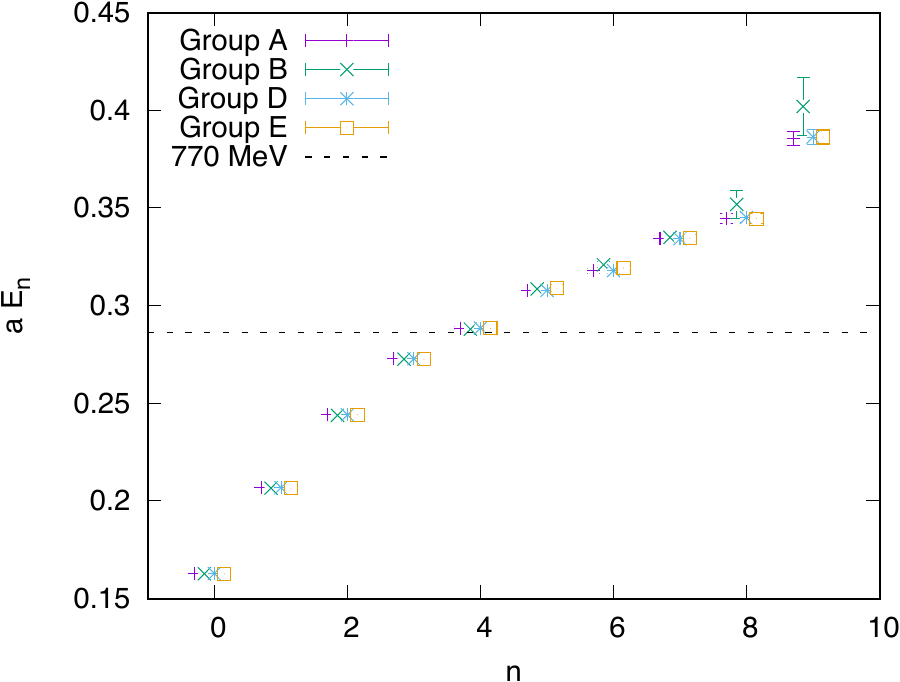}\includegraphics[width=0.48\linewidth]{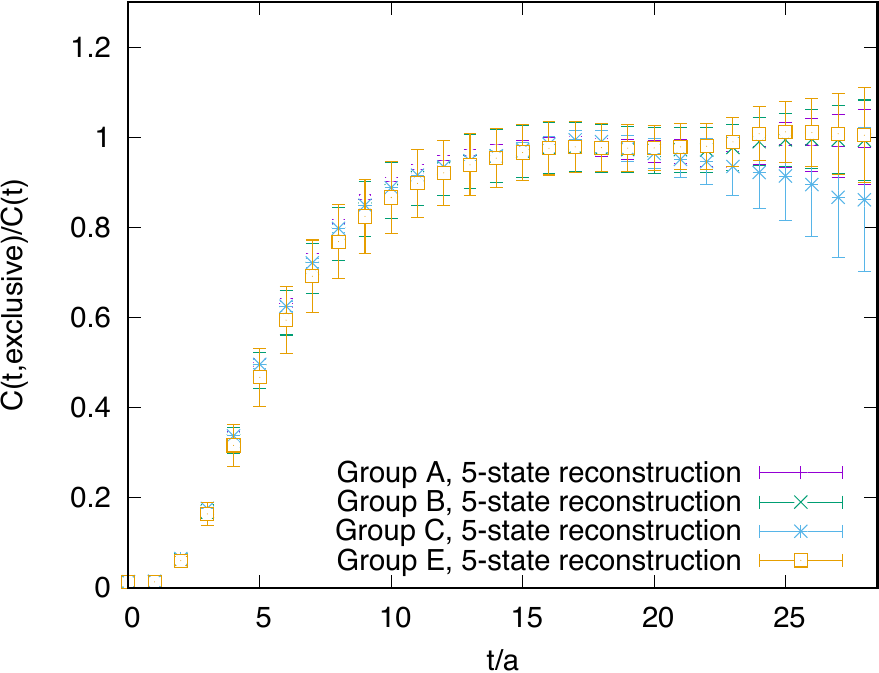}
\caption{\label{fig:checks} Cross-checks of quantities that are not affected by the blinding factor.  In the left panel
we show the check of the spectrum of ensemble 96I and in the right panel we show the ratio of $C^5(t) / C(t)$ for the 64I
ensemble as performed by the different groups.
}
\end{figure*}

In the following, we summarize the main distinct features of the individual
analysis groups that were identified during the relative unblinding:
\begin{itemize}
\item Group A provided a lattice spacing determination that was already used in Ref.~\cite{RBC:2023pvn}.
\item Group B provided an independent lattice spacing determination, which did not include an ansatz
of $w_0 = w_0^{\rm cont} + a^2 w_0^{\rm slope}$ constraining the continuum extrapolation of $w_0$ in the global fit.  This effect is responsible for the larger
statistical noise of group B compared to group A.
\item Groups A and B implement model averaging and compared flat and AIC weights with consistent results.  More details are provided in the following section.
\item Comparing group C and group A, there are three positive $1\sigma$ standard deviation shifts for ensembles 48I, C, and D for group C's analysis compared to group A.
If group A adds these shifts to the analysis, the results of groups A
and C can be brought into exact agreement.  Group C did not propagate
the lattice spacing uncertainty for ensembles C, D, 4, 1, 3, 9, L, which is
the reason for the smaller uncertainties compared to group A.
\item Group D used a third independent lattice spacing determination from a global fit.  The lattice parameters are in good agreement with the results of group A.
\end{itemize}
We also provide an example of the cross-checks that were performed
prior to relative unblinding in Fig.~\ref{fig:checks}.

\section{Model averaging}
\begin{figure}
\includegraphics[width=\linewidth]{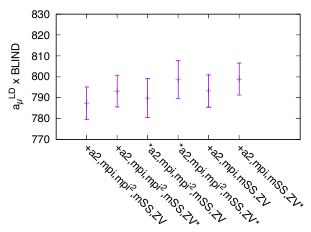}\\
\includegraphics[width=\linewidth]{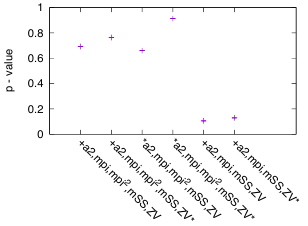}
\caption{\label{fig:fitres} Result of the individual fits of the RBC/UKQCD24 prescription in the BMW20 world.  The
$f_+$ fits are denoted as $+a^2$, the $f_*$ fits are denoted as $*a^2$.  The fits with $f_3=0$ are denoted by the absence of the mpi$^2$ term.
The fits using $Z_V^\pi$ are denoted by ZV and the fits using $Z_V^\star$ are denoted by ZV*.  The top panel shows the
fit results and the bottom panel shows the corresponding p-values.
}
\end{figure}

The individual fit results described in the main text of the manuscript
for the RBC/UKQCD24 prescription are shown in Fig.~\ref{fig:fitres}.
We note that the variation of the fits is of the size of the statistical
uncertainties.  Fits that include a pion mass curvature term are preferred, however,
the p-value of fits using a linear pion mass dependence is not negligible.

\begin{figure}

\includegraphics[width=\linewidth]{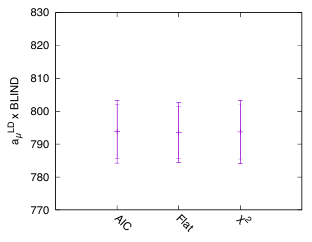}
\caption{\label{fig:ma} We show the final fit results in the RBC/UKQCD24 prescription in the BMW20 world
for different model probabilites $P(M|D)$ in the model averaging procedure.
}
\end{figure}

We perform model averaging by considering a probability $P(M|D)$ of a given
model $M$ given the data $D$ that is either given by the AIC, a $\chi^2$ term only, or a flat weight of constant $P(M|D)$.
For the AIC, we use
\begin{align}
P(M|D) \propto e^{-\frac12\chi^2 - k}
\end{align}
with number of fit parameters $k$ and the proportionality constant chosen such that $\sum_M P(M|D)=1$.  Expectation values of functions $f$ of common
parameters $p$ are defined by
\begin{align}
\langle f(p) \rangle = \sum_M \langle f(p) \rangle_M P(M|D) \,,
\end{align}
where $\langle \cdot \rangle_M$ is the expectation value within a model $M$.
The total variance of $f(p)$ is
\begin{align}
&\langle f(p)^2 \rangle - \langle f(p) \rangle^2\\
&= \sum_M \langle f(p)^2 \rangle_M P(M|D)
-\left(\sum_M \langle f(p) \rangle_M P(M|D)\right)^2 \notag\\
&= \sum_M \sigma^2_{f(p),M} P(M|D) \notag \\ \notag
&+ \sum_M \langle f(p) \rangle_M^2 P(M|D)
-\left(\sum_M \langle f(p) \rangle_M P(M|D)\right)^2
\end{align}
with 
\begin{align}
\sigma^2_{f(p),M} = \langle f(p)^2 \rangle_M - \langle f(p) \rangle_M^2 \,.
\end{align} 
When quoting uncertainties from the model averaging procedure we quote the first line
\begin{align}
  \sigma_{f(p),\rm stat}^2 = \sum_M \sigma^2_{f(p),M} P(M|D)
\end{align}
as the statistical variance and the second line
\begin{align}
  \sigma_{f(p),\rm sys}^2 &= \sum_M \langle f(p) \rangle_M^2 P(M|D) \notag\\&\quad
-\left(\sum_M \langle f(p) \rangle_M P(M|D)\right)^2
\end{align}
as the systematic variance.
By construction the statistical and systematic variance add to the total.  We show in Fig.~\ref{fig:ma} that the results are largely independent of the specific choice of $P(M|D)$.

\clearpage

\end{document}